# p-i-n Tunnel FETs vs. n-i-n MOSFETs: Performance Comparison from Devices to Circuits

Yunfei Gao, Siyuranga O. Koswatta*, Dmitri E. Nikonov[†], and Mark S. Lundstrom
Presenter: Yunfei Gao, Email: gaoy@purdue.edu
Electrical and Computer Engineering, Purdue University, West Lafayette, IN 47906 USA
*IBM T. J. Watson Research Center, Yorktown Heights, NY 10598 USA
† Technology and Manufacturing Group, Intel Corp., Santa Clara, CA 95052 USA

## ABSTRACT

The band-to-band tunneling transistors have some performance advantages over the conventional MOSFETs due to the <60mV/dec sub-threshold slope. In this paper, carbon nanotubes are used as a model channel material to address issues that we believe will apply to BTBT FETs vs. MOSFETs more generally. We use $p_z$-orbital tight-binding Hamiltonian and the non-equilibrium Green's function (NEGF) formalism for rigorous treatment of dissipative quantum transport [3]. A device level comparison of p-i-n TFETs and n-i-n MOSFETs in both ballistic and dissipative cases has been performed previously [2]. In this paper, the possibility of using p-i-n TFETs in ultra-low power sub-threshold logic circuits is investigated using a rigorous numerical simulator [4]. The results show that, *in sub-threshold circuit operation*, the p-i-n TFETs have *better DC characteristics*, and *can deliver ~15x higher performance at the iso-$P_{LEAKAGE}$, iso-$V_{DD}$ conditions*. Because p-i-n TFETs can operate at lower $V_{DD}$ than n-i-n MOSFETs, they *can deliver ~3x higher performance at the same power ($P_{OPERATION}$)*. This results in *~3x energy reduction under iso-delay conditions*. Therefore the p-i-n TFETs are more suitable for sub-threshold logic operation.

## INTRODUCTION

Band-to-band tunneling (BTBT) transistors based on the p-i-n geometry have been extensively studied recently [1]. Compared to the n-i-n MOSFET geometry, the p-i-n tunnel FETs (TFETs) can produce a sharper sub-threshold slope below the MOSFET limit of 60mV/decade at room temperature. This leads to lower off-current ($I_{OFF}$) and thus smaller leakage power ($P_{LEAKAGE}$). However, its on-current ($I_{ON}$) is limited due to the existence of the tunneling barrier. Phonon scattering can also limit the performance of p-i-n TFETs in the off-state [2]. Due to the different pros and cons of p-i-n TFETs over n-i-n MOSFETs in intrinsic device operation, we did a systematic evaluation of these two types of devices from basic physics to circuit performance, and conclude that the p-i-n TFETs have performance advantage in sub-threshold logic.

## SIMULATION METHODOLOGY

The device structure employed in this study is a (13, 0) carbon nanotube with a cylindrical high-κ (κ=16, $t_{OX}$=2nm) gate geometry and 15nm channel length. The drain region is doped n-type, and the source is doped n-type or p-type to produce n-i-n MOSFET or p-i-n TFET, respectively. In the NEGF formalism phonon scattering is treated by a self-energy function for electron-phonon interaction [3]. A full spectrum of $I_{DS}$ values is obtained at different $V_{GS}$ and $V_{DS}$ varying from 0 to $V_{DD}$ (2D I-V). These results are then used in the circuit simulation. Another input is the 2D $C_G$-$V_{GS}$,$V_{DS}$ data, which we calculate numerically from the charge induced in the device. Then we simplify it into 1D $C_G$-$V_{GS}$ dependence with a reasonably general relation between $V_{GS}$ and $V_{DS}$ in circuits [5]. After obtaining the 2D I-V and 1D C-V data, voltage domain simulation is used to calculate DC and transient characteristics (delay, power consumption and energy) of various kinds of benchmark circuits [4]. According to the projections for a similar size Si CMOS, the parasitic capacitance is comparable to intrinsic capacitance [6]. We take a similar ratio here to capture parasitic effects. We shift the flat band voltages for the p-i-n TFET and n-i-n MOSFET to operate at the same $I_{OFF}$. Under this condition, leakage power is nearly the same (iso-$P_{LEAKAGE}$). We focus on sub-threshold logic characteristics by keeping $V_{DD}$ below $V_T$ of both devices. Note that for each $V_{DD}$, the flat band voltage needs to be adjusted to obtain the iso-$I_{OFF}$ condition.

## SIMULATION RESULTS

We include *phonon scattering* in all the simulations unless otherwise stated. Fig. 1b shows ~40mV/decade sub-threshold slope for a p-i-n TFET. At high $V_{GS}$, its $I_{ON}$ is smaller than that in a MOSFET due to the presence of a tunnel junction between the source and the channel. The effect of phonon scattering and temperature can also be seen in Fig. 1.

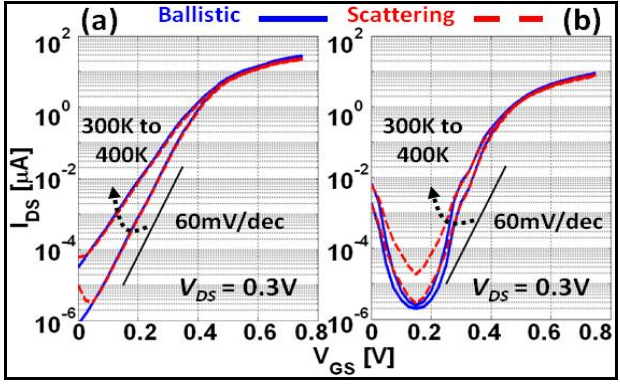

Fig. 1. $I_{DS}$-$V_{GS}$ dependence on temperature and phonon scattering for (a) n-i-n MOSFET and, (b) p-i-n TFET. The latter has reduced temperature dependence under ballistic conditions because of the energy-filtering tunneling mechanism at the source-channel junction. Phonon scattering can, however, degrade the off-state performance of the p-i-n TFET due to phonon absorption assisted transport. Higher temperature degrades the off-state in n-i-n MOSFETs since more electrons can participate in thermionic emission transport over the channel barrier [2].

Note that in p-i-n TFETs, in order to get rid of the >60mV/dec swing region in the off-state (Fig. 2a), we should engineer contact doping, bandgap, and gate work-function to suppress hole conduction, and to align channel $E_C$ and source $E_V$ at around $V_{GS}$=0.

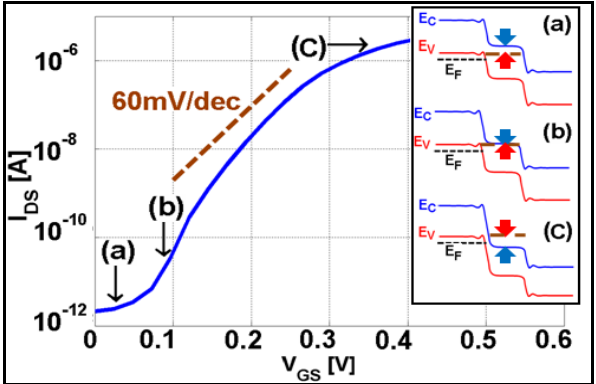

Fig. 2. Evolution of band diagrams (inset) corresponding to different values $V_{GS}$ (in the plot). $V_{FLATBAND}$ is shifted so the minimum current occurs at $V_{GS}$=0. The channel $E_C$ is above the source $E_V$ (inset a), which gives >60mV/dec. With gradually raising $V_{GS}$, channel $E_C$ abruptly drops below source $E_V$, leading to <60mV/dec region (inset b). Increasing the gate bias further (inset c), sub-threshold slope starts to degrade again. It is preferable to adjust the band gap and source drain doping to eliminate region (a), and start from (b) at $V_{GS}$=0V.

The gate capacitance strongly depends on $V_{DS}$ in p-i-n TFETs than in n-i-n MOSFETs [2]. From Figs. 3 and 4 we see that $C_{GS}$ and $C_{GD}$ are equal in n-i-n MOSFETs, whereas $C_{GD}$ dominates in p-i-n TFETs.

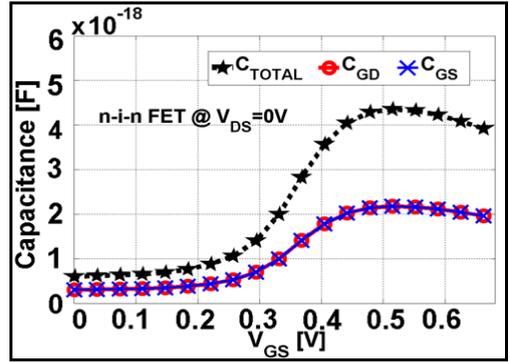

Fig. 3. $C_{GS}$,$C_{GD}$ vs. $V_{GS}$ at $V_{DS}$=0V for a *dissipative* n-i-n MOSFET. $C_{GS}$ and $C_{GD}$ represent the states in the channel filled by the source and the drain, respectively. In the n-i-n MOSFET, the total capacitance is equally divided into $C_{GS}$ and $C_{GD}$, which is due to the symmetry of the source and drain regions.

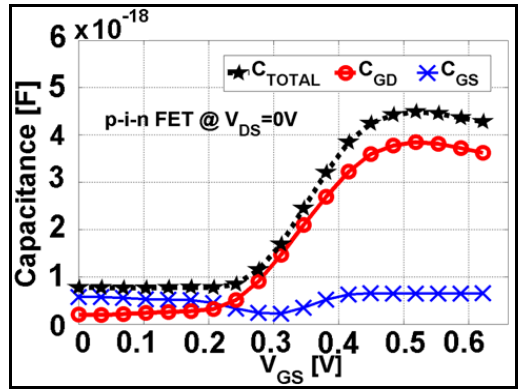

Fig. 4. $C_{GS}$,$C_{GD}$ vs. $V_{GS}$ at $V_{DS}$=0V for a *dissipative* p-i-n MOSFET. $C_{GS}$ and $C_{GD}$ represent the states in the channel filled by the source and the drain, respectively. In the p-i-n TFET, $C_{GD}$ is much larger than $C_{GS}$, because the tunneling barrier between source and channel forbids the source electrons from entering the channel. This effect makes the controlling of $C_{GD}$ more important in p-i-n TFET than in n-i-n MOSFET.

The dominance of $C_{GD}$ is explained in another way in Fig. 5. It's because the source-evolving local density of states (LDOS) in the channel is very small compared to drain-evolving LDOS [2]. This makes the gate-to-drain quantum capacitance large.

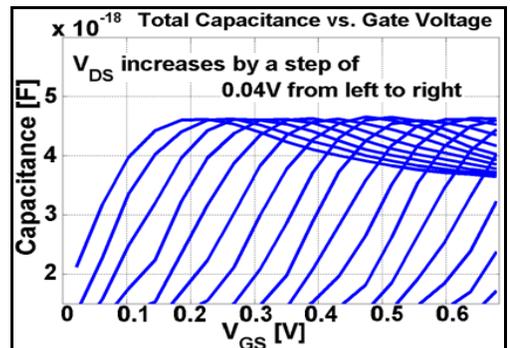

Fig. 5. Capacitance-$V_{GS}$ dependence for different $V_{DS}$ in *ballistic* p-i-n TFETs. It shows that for every increase of 0.04V in the drain bias, the capacitance would remain the same if the gate bias is increased by the similar amount. In other words, as the relative position of energy bands in the channel and drain remain unchanged, the total capacitance is fixed. This indicates that most of the device capacitance is contributed by the drain. This dependence can also be observed in *dissipative* simulation.

All circuit simulations are done at the iso-$P_{LEAKAGE}$ condition. It allows a fair comparison of delay and dynamic power for both devices. Fig. 6 shows that the *ballistic* p-i-n TFET has near-ideal voltage transfer characteristics (VTC), and it is degraded by phonon scattering. The n-i-n MOSFET VTC shows an even lower noise margin.

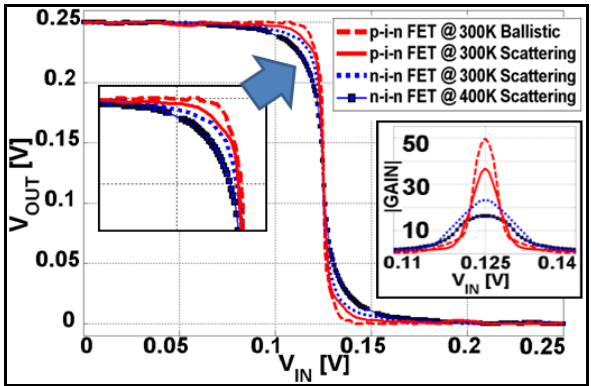

Fig. 6. The voltage transfer characteristic (VTC) of a n-i-n MOSFET and p-i-n TFET in the sub-threshold region at two temperatures, with and without phonon scattering. The *ballistic* p-i-n TFET shows a nearly-ideal VTC at both 300K and 400K (not shown), but phonon scattering affects the VTC at these two temperatures (the 400K case with scattering is not shown). The *dissipative* n-i-n MOSFET shows degradation at higher temperatures, while the ballistic operation has little effect on VTC (not shown). The inverter gains are plotted in the inset.

Because the two devices have similar $P_{LEAKAGE}$ and the same $V_{DD}$, the p-i-n TFET can deliver ~15x higher performance than the n-i-n MOSFET in simple benchmark circuits such as an inverter driving another identical inverter (Fig. 7) and a 10-stage 1-fanout NAND chain (Fig. 8).

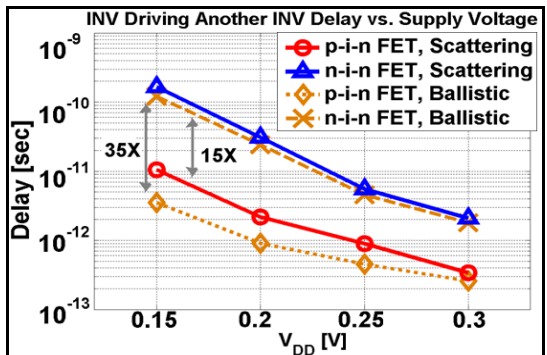

Fig. 7. Delay vs. $V_{DD}$ for an inverter driving another identical inverter *with phonon scattering*. The dotted line shows the corresponding delay in *ballistic* case. At iso-leakage condition, the *dissipative* p-i-n TFET circuit performance can be ~15x better than that of the n-i-n MOSFET. Increasing $V_{DD}$ can improve the performance of both devices, but the operation power will increase.

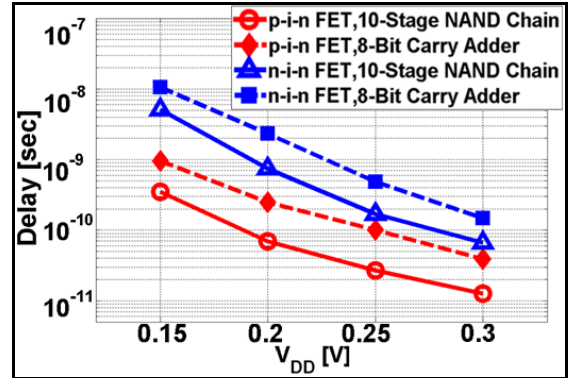

Fig. 8. Delay vs. $V_{DD}$ of a 10-stage 1-fanout NAND chain and an 8-Bit Carry Adder. Under *dissipative* conditions, p-i-n TFETs have a performance advantage over n-i-n MOSFETs in the sub-threshold region. This holds not only for the simple benchmark circuit such as inverter (Fig. 8) and NAND gate, but also in larger logic circuits.

When considering larger logic circuits such as an 8-bit carry adder, the p-i-n TFET has ~10x higher performance (Fig 8). As shown in Fig. 9, we can operate the p-i-n TFET at a lower $V_{DD}$ to reduce its operation power ($P_{OPERATION}$), while the advantage in performance is still retained.

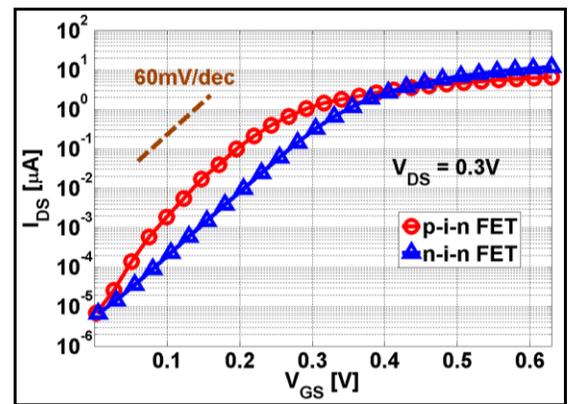

Fig. 9. $I_{DS}$-$V_{GS}$ dependence of *dissipative* n-i-n MOSFET and p-i-n TFET. The two curves are shifted to match $I_{OFF}$. In this case $V_T$~0.35V, below which is the region of interest for sub-threshold circuit operation. In that region the p-i-n TFET can deliver the same amount of current at a lower $V_{DD}$ as the n-i-n MOSFET which makes the p-i-n TFETs suitable for ultra-low power application with a moderate frequency of operation.

In the inverter driving another identical inverter case (Fig. 10), the p-i-n TFET can obtain the same performance as the n-i-n MOSFET at a lower $V_{DD}$.

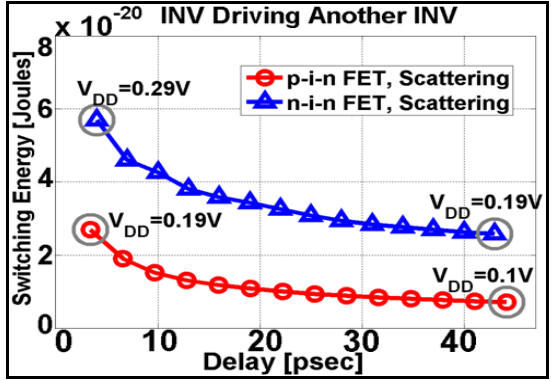

Fig. 10. Switching energy vs. delay for an inverter driving another identical inverter *with phonon scattering*. At the same delay, the n-i-n MOSFET consumes ~3x more energy than the p-i-n TFET. With the same amount of energy, the p-i-n TFET has faster operation. Note that the supply voltage for the p-i-n TFET is smaller than that for the n-i-n MOSFET. With this smaller $V_{DD}$, the p-i-n TFET can still reach high enough current to ensure a shorter delay for sub-threshold operation.

This results in ~3x smaller switching energy and ~3x smaller $P_{OPERATION}$ for the p-i-n TFET since $E_{SWITCH}=C_{SWITCH}V_{DD}^2$, and $C_{SWITCH}$ is nearly the same in these two devices and remains almost constant for different $V_{DD}$ (Fig. 11b). Note that since the $I_{OFF}$ are almost the same in these two devices, the leakage powers are very similar as shown in Fig. 11a.

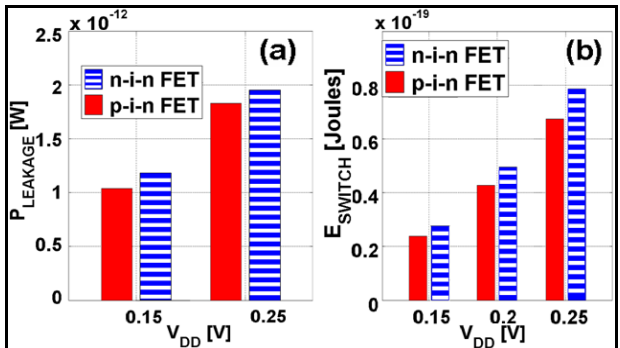

Fig. 11. (a) The leakage power ($P_{LEAKAGE}$) vs. $V_{DD}$ for an inverter driving another identical inverter *with phonon scattering*. Since $I_{OFF}$ of n-i-n MOSFET and p-i-n TFET are the same, the $P_{LEAKAGE}$ is almost the same, with little difference mainly due to the DIBL effect [2]. Same iso-$P_{LEAKAGE}$ trend exists in other benchmark circuits. (b) Switching energy ($E_{SWITCH}$) vs. $V_{DD}$ for the same case. The quadratic relation, $E_{SWITCH}=C_{SWITCH}V_{DD}^2$ is clearly seen. $C_{SWITCH}$ of p-i-n TFET is smaller than that of n-i-n MOSFET.

When an inverter drives a large constant load (Fig. 12), the p-i-n TFET can deliver ~2x higher performance with the same $P_{OPERATION}$, which in turn translates into ~3x lower $P_{OPERATION}$ at the iso-delay condition.

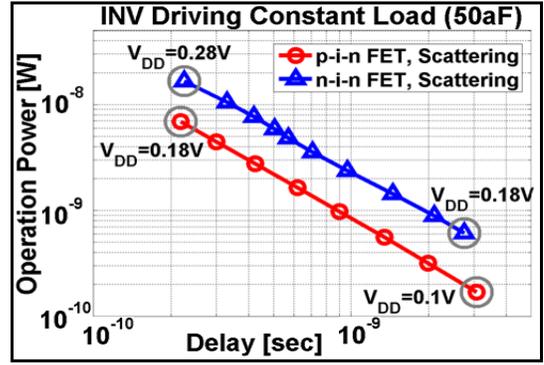

Fig. 12. Delay vs. operation power for the inverter driving a constant load of 50aF. All the simulations are in the sub-threshold region and include *phonon scattering*. Under iso-leakage, iso-delay condition, the n-i-n MOSFET consumes ~3x more power than p-i-n TFET; while under iso-leakage, iso-power condition, n-i-n MOSFET has a ~2x larger delay than p-i-n TFET. This means that in the sub-threshold region the p-i-n TFET can deliver moderate performance with much lower power consumption.

In summary, we make a comparison between *dissipative* n-i-n MOSFETs and p-i-n TFETs from device physics to circuit performance. The potential of applying p-i-n TFETs in sub-threshold logic circuits is explored. Better DC characteristics and a ~15x higher performance (iso-$P_{LEAKAGE}$) are seen in *dissipative* p-i-n TFETs. Also it consumes ~3x less energy for the same performance as n-i-n MOSFETs. We see that even with phonon scattering and parasitic effect, the p-i-n TFETs still have advantage over the n-i-n MOSFETs in sub-threshold logic circuit operation.